\documentstyle[12pt,YN,epsbox]{article}
\newenvironment{subequations}{
    \begingroup

    \setcounter{enumi}{\value{equation}}
    \refstepcounter{enumi}
    \setcounter{equation}{0}
}{
    \setcounter{equation}{\value{enumi}}
    \endgroup
    \par
}

\newcommand{\refsec}[1]{Sec.~\ref{#1}}
\newcommand{\refref}[1]{\cite{#1}}
\newcommand{\reffig}[1]{Fig.~\ref{#1}}

\newcommand{\ds}{\displaystyle}
\newcommand{\seq}[1]{\mbox{\scriptsize #1}}
\newcommand{\teq}[1]{\mbox{\tiny #1}}

\newcommand{\p}{\partial}
\newcommand{\pdif}[1]{\frac{\p}{\p #1}}
\newcommand{\fdif}[1]{\frac{\delta}{\delta #1}}
\newcommand{\FDIF}[2]{\frac{\delta #1}{\delta #2}}
\newcommand{\pint}[1]{\int{\cal D}#1}
\newcommand{\bpint}[3]{\int^{#1}_{#2}{\cal D}#3}
\newcommand{\la}{\langle}
\newcommand{\ra}{\rangle}
\newcommand{\w}{\omega}
\newcommand{\T}{\mbox{T}}

\newcommand{\lb}{\la}
\newcommand{\rb}{\ra_{\seq{bc}}}

\newcommand{\tI}{t_{\seq{I}}}
\newcommand{\tF}{t_{\seq{F}}}
\newcommand{\xI}{x_{\seq{I}}}
\newcommand{\xF}{x_{\seq{F}}}
\newcommand{\sti}{t_{\teq{I}}}
\newcommand{\stf}{t_{\teq{F}}}
\newcommand{\sxi}{x_{\teq{I}}}
\newcommand{\sxf}{x_{\teq{F}}}

\newcommand{\xf}{x^{(0)}}
\newcommand{\xclf}{x_{\seq{cl}}^{(0)}}
\newcommand{\xQf}{x_{\seq{Q}}^{(0)}}
\newcommand{\Gf}{G^{(0)}_{\seq{bc}}}
\newcommand{\Df}{D^{(0)}_{\seq{bc}}}
\newcommand{\prop}{\Delta_{\seq{bc}}}

\begin{document}
\begin{flushright}
%\today \\
WU-HEP-96-12\\
\end{flushright}
\vspace{.5cm}
\begin{center}
\Large{\bf
Transition Amplitude within the Stochastic Quantization
Scheme}\\
\large
-- Perturbative Treatment --\\
\vspace{16pt}
\newcounter{Footnote}
\renewcommand{\thefootnote}{\fnsymbol{Footnote}}
Kazuya Yuasa$^{*}$\footnote{%
$^{*}\,$email address: 696L1136@mn.waseda.ac.jp
}
and
Hiromichi Nakazato$^{**}$\footnote{%
$^{**}\,$email address: hiromici@mn.waseda.ac.jp
}\\
\renewcommand{\thefootnote}{\arabic{footnote}}
\setcounter{footnote}{0}
\vspace{12pt}
{\it
Department of Physics, Waseda University, Tokyo 169, Japan
} \\

%\vspace*{.5cm} PACS:
\vspace*{1.5cm}

{\small\bf Abstract}\\ \end{center}

{\small
The quantum mechanical transition amplitudes are calculated
perturbatively on the basis of the stochastic quantization
method of Parisi and Wu.
It is shown that the stochastic scheme reproduces the
ordinary result for the amplitude and systematically
incorporates higher-order effects, even at the lowest order.
}
\newpage

\section{Introduction}
\setcounter{equation}{0}
\label{sec:intro}
Since its proposal, the stochastic quantization method of
Parisi and Wu\cite{ref:P-W} has been widely applied to
various fields in physics\cite{ref:sqm-rev}, quite
straightforwardly at relatively early stages of its
development and with some conceptual and/or technical
innovation at later times.
A scheme for obtaining the quantum mechanical transition
amplitude on the basis of the stochastic quantization (SQ)
was proposed by H.\ H\"uffel and one of the present authors
a few years ago\cite{ref:hh-hn} and is considered to belong
to the latter type of applications:
One may easily realize the difficulty in obtaining an {\it
amplitude} from a {\it probability}, on which the stochastic
process, even if fictitious in this scheme, is essentially
based.
In this paper, this scheme\cite{ref:hh-hn} is applied to a
nonlinear system, for which no exact expression of the
transition amplitude is known and we have to rely on a
perturbative treatment.
The purpose of the paper is to show how a perturbative
treatment is possible within this scheme, to confirm its
consistency with the ordinary one and then to find possible
advantages over the latter.

After brief reviews of SQ in \refsec{sec:SQM} and of the
stochastic scheme for the transition amplitude in
\refref{ref:hh-hn} in \refsec{sec:formula}, we describe how
to treat nonlinear systems within this scheme and propose a
perturbative treatment in \refsec{sec:formal}.
Section \ref{sec:example} demonstrates some details of the
calculation of the transition amplitudes for a quantum
mechanical anharmonic oscillator.
It will be clear that our treatment is equivalent to a
perturbative expansion of the logarithm of the amplitude and
therefore the amplitude thus obtained includes automatically
a part of the higher-order contributions as an exponential
form.
The last section, \refsec{sec:summary}, is devoted to a
summary.

\section{Review of the Stochastic Quantization Method}
\setcounter{equation}{0}
\label{sec:SQM}
Let us briefly outline the ordinary prescription of the
stochastic quantization method of Parisi and
Wu\cite{ref:P-W,ref:sqm-rev}.
This quantization method is so designed as to give quantum
mechanics as the thermal-equilibrium limit of a hypothetical
stochastic process.
For this purpose, the dynamical variable $x(t)$\footnote{For
simplicity, we exclusively consider one-dimensional quantum
mechanical systems here.} is assumed to be a stochastic
variable $x(t,s)$ with respect to a newly-introduced time,
called the ``fictitious time" $s$.
Its dynamics is given by the Langevin equation\footnote{We
work in the Minkowski stochastic quantization
scheme\cite{ref:Minkowski_SQM} throughout the paper.}
\begin{equation}
\frac{\p}{\p s}x(t,s) = i\FDIF{S}{x(t,s)} + \eta(t,s),
\label{eqn:Langevin_eq}
\end{equation}
where $S$ is the classical action of the system and $\eta$
is the Gaussian white noise, whose statistical properties
are characterized by
\begin{equation}
\la\eta(t,s)\ra = 0, \;\;\;
\la\eta(t,s)\eta(t',s')\ra = 2\delta(t-t')\delta(s-s'),
\;\;\; \mbox{\it etc.}
\label{eqn:noise}
\end{equation}
We solve the Langevin equation (\ref{eqn:Langevin_eq}) under
some initial condition to get $x(t,s)$ as a functional of
the noise $\eta$, calculate the equal-time correlation
function $\la x(t_1,s)x(t_2,s)\cdots\ra$ by means of
(\ref{eqn:noise}), and take the equilibrium limit
$s\to\infty$.
What we thereby obtain is shown to correspond to the
ordinary vacuum expectation value $\la0|\T
x(t_1)x(t_2)\cdots|0\ra$.

This correspondence is most clearly seen in the
Fokker-Planck picture.
In this picture the stochastic average $\la\cdots\ra$ is
expressed as a functional integral
\begin{equation}
\la x(t_1,s)x(t_2,s)\cdots\ra
=\pint{x}\,x(t_1)x(t_2)\cdots P[x;s]
\label{eqn:expectation_FP}
\end{equation}
with a (quasi-)probability distribution functional $P$ which
obeys the Fokker-Planck equation
\begin{equation}
\frac{\p}{\p s}P[x;s]
=\int_{-\infty}^{\infty}dt
  \fdif{x(t)}\left(\fdif{x(t)}-i\FDIF{S}{x(t)}\right)P[x;s]
\label{eqn:FokkerPlanck_eq}
\end{equation}
and is normalized as $\la1\ra=\int{\cal D}xP[x;s]=1$.
Clearly, the stationary solution of this equation is given
by $e^{iS}$, which also serves as the equilibrium
distribution if we adopt the so-called
$i\varepsilon$-prescription\cite{ref:Minkowski_SQM}.
Under this prescription, the equal-time correlation function
(\ref{eqn:expectation_FP}) approaches, as
$s\rightarrow\infty$, the Feynman path integral, which is
nothing but the vacuum expectation value in the canonical
(operator) formalism:
\begin{equation}
\lim_{s\rightarrow\infty}\la x(t_1,s)x(t_2,s)\cdots\ra
=\frac{\ds\pint{x}\,x(t_1)x(t_2)\cdots e^{iS}}
      {\ds\pint{x}\,e^{iS}}
= \la0|\T x(t_1)x(t_2)\cdots|0\ra.
\label{eqn:expectation_L}
\end{equation}

\section{The Stochastic Formula for Transition Amplitude}
\setcounter{equation}{0}
\label{sec:formula}
The transition amplitude is one of the fundamental
quantities in quantum mechanics, for the amplitude plays a
central role there.
On the other hand, as one readily notices, the above
stochastic scheme apparently provides us with those
expectation values that are normalized in the sense of
(\ref{eqn:expectation_L}).
This can be thought of as rooted in the essential property
of the stochastic process, where the probability (density),
not the amplitude, governs the dynamics.
Therefore it does not seem trivial to derive amplitudes
within the framework of the stochastic formalism of Parisi
and Wu.

Let us first try to calculate the correlation function,
under the boundary conditions
\begin{equation}
x(\tF,s) = \xF, \;\;\; x(\tI,s) = \xI.
\label{eqn:boundary}
\end{equation}
These conditions are considered to be one of the simplest
extensions of those for the quantum mechanical transition
amplitude $\la\xF,\tF|\xI,\tI\ra$.
It would be natural to expect that the correlation function
under the above boundary conditions $\lb
x(t_1,s)x(t_2,s)\cdots\rb$ is expressed in the Fokker-Planck
picture as
\begin{equation}
\lb x(t_1,s)x(t_2,s)\cdots\rb
=\bpint{x(\stf)=\sxf}{x(\sti)=\sxi}{x}\,
                x(t_1)x(t_2)\cdots P[x;s]
\label{eqn:bexpectation_FP}
\end{equation}
with the normalization condition $\lb1\rb=1$.
The quantity with the subscript ${}_{\rm bc}$ is to be
evaluated within SQ under the above boundary conditions
(\ref{eqn:boundary}).
This normalization condition implies that in the
$s\rightarrow\infty$ limit, the correlation function
approaches the following normalized quantity
\begin{eqnarray}
\lim_{s\rightarrow\infty}\lb x(t_1,s)x(t_2,s)\cdots\rb
&=&\frac{
\ds\bpint{x(\stf)=\sxf}{x(\sti)=\sxi}{x}\,
                x(t_1)x(t_2)\cdots e^{iS}
}{
\ds\bpint{x(\stf)=\sxf}{x(\sti)=\sxi}{x}\,e^{iS}
}\nonumber\\
\noalign{\vspace{6pt}}
&=& \frac{\la\xF,\tF| \T x(t_1)x(t_2)\cdots |\xI,\tI\ra}
         {\la\xF,\tF|\xI,\tI\ra}.
\label{eqn:bexpectation_L}
\end{eqnarray}
It is such normalized expectation values
(\ref{eqn:bexpectation_L}) that we can directly calculate
within this scheme.
Remember that what we are interested in here is not the
vacuum expectation value, but the transition amplitude
$\la\xF,\tF|\xI,\tI\ra$ itself or, in other words, the
normalization factor ({\it i.e.}\ the denominator in
(\ref{eqn:bexpectation_L})) itself.

In \refref{ref:hh-hn}, a formula for the transition
amplitude within the framework of SQ is proposed and applied
to several solvable cases.
The formula is based on the following relations in quantum
mechanics
\begin{subequations}
  \begin{equation}
  -i\frac{\p}{\p\tI}\la\xF,\tF|\xI,\tI\ra
  = \la\xF,\tF|H(\tI)|\xI,\tI\ra
  \label{eqn:t_eq}
  \end{equation}
and
  \begin{equation}
  i\frac{\p}{\p\xI}\la\xF,\tF|\xI,\tI\ra
  = \la\xF,\tF|p(\tI)|\xI,\tI\ra,
  \label{eqn:x_eq}
  \end{equation}
\end{subequations}\noindent
where $H(\tI)$ and $p(\tI)$ are the Hamiltonian and the
momentum operators, respectively.
{}From (\ref{eqn:t_eq}) we get
\begin{equation}
\frac{\p}{\p\tI}\ln \la\xF,\tF|\xI,\tI\ra
= i\frac{\la\xF,\tF|H(\tI)|\xI,\tI\ra}
        {\la\xF,\tF|\xI,\tI\ra}
\equiv i\lb H(\tI)\rb,
\end{equation}
that has the solution
\begin{subequations}
\label{eqn:formulae}
    \begin{equation}
    \la\xF,\tF|\xI,\tI\ra
    = c\exp\left[i\int^{\sti}d\tI\lb H(\tI)\rb\right],
    \label{eqn:toHamiltonian}
    \end{equation}
with a factor $c$ dependent on $\xF$, $\xI$ and $\tF$.
Along the same line of thought, we obtain from
(\ref{eqn:x_eq})
    \begin{equation}
    \la\xF,\tF|\xI,\tI\ra
    = \tilde{c}
      \exp\left[-i\int^{\sxi}d\xI\lb p(\tI)\rb\right]
    \label{eqn:toMomentum}
    \end{equation}
\end{subequations}\noindent
with a factor $\tilde{c}$ dependent on $\xF$, $\tF$ and
$\tI$.
The transition amplitude $\la\xF,\tF|\xI,\tI\ra$ has thus
been related to the (normalized) expectation value of the
Hamiltonian $\lb H(\tI)\rb$ and to that of the momentum $\lb
p(\tI)\rb$, which are obtainable as the equilibrium limits
of $ \lb H(\tI,s)\rb$ and $\lb p(\tI,s)\rb$ in
SQ.\footnote{Equation (\ref{eqn:toHamiltonian}) is the
formula presented in \refref{ref:hh-hn}.
On the other hand, though it is not explicitly presented
there, (\ref{eqn:toMomentum}) is surely used in the
practical calculations.}

Notice that these relations are nothing but two of the
expressions of the ordinary variational principle and there
are two more formulae, {\it i.e.}, the formulae relating the
transition amplitude to $\lb H(\tF)\rb$ and $\lb p(\tF)\rb$.
For most systems, however, the above two formulae
(\ref{eqn:formulae}) are sufficient to determine $c$ and
$\tilde{c}$ except for a constant factor independent of
$\tF$, $\tI$, $\xF$ and $\xI$.
In fact, we can determine the $\xI$-dependence of $c$ from
(\ref{eqn:toMomentum}) and the $\tI$-dependence of
$\tilde{c}$ from (\ref{eqn:toHamiltonian}).
Furthermore, their $\xF$- and $\tF$-dependences are fixed
from the fact that $\la\xF,\tF|\xI,\tI\ra$ is a function of
$T\equiv\tF-\tI$ (translational invariance) and is a
symmetric function of $\xF$ and $\xI$ (time-reversal
invariance).
The constants $c$ and $\tilde{c}$ are thus completely
determined if we fix the remaining factor by requiring
$\la\xF,\tF|\xI,\tI\ra$ to approach a Dirac
$\delta$-function $\delta(\xF-\xI)$ as $T\rightarrow0$.

\section{A Perturbative Treatment}
\setcounter{equation}{0}
\label{sec:formal}
The above formulae (\ref{eqn:formulae}) have been applied in
\refref{ref:hh-hn} to linear systems to derive transition
amplitudes within the framework of SQ.
It is demonstrated that the stochastic scheme can reproduce
the correct results for the amplitudes.
In the following, we shall develop a perturbative method in
order to treat nonlinear systems within this scheme.

Since all that is needed in deriving the amplitude in SQ are
the expectation values of $H$ and $p$, according to
(\ref{eqn:toHamiltonian}) and (\ref{eqn:toMomentum}), a
natural strategy would be to evaluate these quantities as a
power series in the coupling constant.
For systems in which the Hamiltonian $H$ is composed of a
kinetic part $p^2/2M$ and a potential part $V(x)$, what
essentially remains to be evaluated is the expectation value
of the former, owing to the boundary conditions
(\ref{eqn:boundary})
\begin{equation}
\lb H(\tI)\rb={1\over2M}\lb p^2(\tI)\rb+V(\xI).
\label{eqn:hbc}
\end{equation}
We have to perturbatively evaluate the expectation values of
momentum and squared momentum.
Though this can be done within the framework of the phase
space formalism of SQ\cite{ref:ps-sq,ref:hh-hn}, we prefer
to work in ordinary configuration space for simplicity.
It is not difficult to derive the following relations
\begin{subequations}
\label{eqn:p_configuration}
    \begin{equation}
    \lb p(\tI)\rb=M\lim_{t\rightarrow\sti}
            \p_{t}\lim_{s\rightarrow\infty}\lb x(t,s)\rb
    \label{eqn:pti}
    \end{equation}
and
    \begin{equation}
    \lb p^2(\tI)\rb
    = M^2\lim_{t_{1},t_{2}\rightarrow\sti}
     \p_{t_{1}}\p_{t_{2}}\lim_{s\rightarrow\infty}
       \lb x(t_{1},s)x(t_{2},s)\rb,
    \label{eqn:ppti}
    \end{equation}
\end{subequations}\noindent
on the basis of the phase-space Langevin
equations.\footnote{The second relation just corresponds to
Feynman's time-splitting procedure in evaluating the average
square velocity\cite{ref:FH}.}
Thus the problem is reduced to the evaluation of the
correlation functions, $\lb x(t,s)\rb$ and $\lb
x(t_{1},s)x(t_{2},s)\rb$:
We solve the Langevin equation (\ref{eqn:Langevin_eq})
perturbatively and get its solution $x(t,s)$ as a power
series in the coupling constant.
As in the solvable cases treated in \refref{ref:hh-hn}, the
boundary conditions (\ref{eqn:boundary}) shall be taken into
account by the classical solution and the remaining part (to
be quantized in terms of the Langevin equation) is subject
to the ``zero boundary conditions" and is expressed as a
Fourier series.

It is worth mentioning here that the perturbative treatment
on the basis of the formulae (\ref{eqn:toHamiltonian}) and
(\ref{eqn:toMomentum}) differs from the ordinary one:
What is really evaluated perturbatively is not the amplitude
itself, as in the ordinary case, but its logarithm.
Therefore we can expect that even the lowest-order value of
$\lb H(\tI)\rb$ or $\lb p(\tI)\rb$ will systematically yield
higher-order contributions to the amplitude.

\section{An Example: Anharmonic Oscillator}
\setcounter{equation}{0}
\label{sec:example}
Let us consider an anharmonic oscillator described by the
action
\begin{equation}
S=\int^{\stf}_{\sti}dt\left(
 \frac{1}{2}M\dot{x}^2-\frac{1}{2}M\omega^2x^2-\frac{g}{4}x^4
\right)
\label{eqn:model}
\end{equation}
and calculate the transition amplitude
$\la\xF,\tF|\xI,\tI\ra$ according to the above stochastic
scheme.
The Langevin equation (\ref{eqn:Langevin_eq}) reads
\begin{equation}
\p_s x(t,s)=-iM(\p_t^2+\w^2)x(t,s)-igx^3(t,s)+\eta(t,s)
\label{eqn:Langevin_model}
\end{equation}
and we write its solution as the sum of the solution for a
harmonic oscillator ($g=0$) $\xf$ and a remaining part
\begin{equation}
x(t,s)=\xf(t,s)+x'(t,s).
\label{eqn:x=x0+x'}
\end{equation}
The free solution $\xf$ is composed of two terms
\begin{equation}
\xf(t,s)=\xclf(t)+\xQf(t,s),
\label{eqn:x0}
\end{equation}
where $\xclf$ is the solution of the classical equation of
motion
\begin{equation}
(\p_t^2+\w^2)\xclf(t)=0\quad\mbox{with}\quad
\xclf(\tF)=\xF,\;\;\;\xclf(\tI)=\xI
\label{eqn:xcl}
\end{equation}
and $\xQf$ that of a free Langevin equation
\begin{equation}
\p_s\xQf(t,s)=-iM(\p_t^2+\w^2)\xQf(t,s)+\eta(t,s)
\quad\mbox{with}\quad\xQf(\tF)=\xQf(\tI)=0,
\label{eqn:xQ}
\end{equation}
where the last conditions are called ``zero boundary
conditions."
They are readily calculated to be
\begin{equation}
\xclf(t)
=\frac{1}{\sin\w T}[\xF\sin\w(t-\tI)+\xI\sin\w(\tF-t)]
\label{eqn:xclf}
\end{equation}
and
\begin{equation}
\xQf(t,s)
=\int^{\stf}_{\sti}dt'\int^{\infty}_{-\infty}ds'
  \Gf(t,t';s-s')\eta(t',s'),
\label{eqn:xQf}
\end{equation}
where
\begin{equation}
\Gf(t,t';s-s')
= \theta(s-s')
    \frac{2}{T}\sum_{n=1}^{\infty}
      \sin\frac{n\pi}{T}(t-\tI)\sin\frac{n\pi}{T}(t'-\tI)
      e^{-iM\w^2\left[
                  1-\left(\frac{n\pi}{\w T}\right)^2
                \right](s-s')}
\label{eqn:Gf}
\end{equation}
is the free Green function which respects the ``zero
boundary conditions" (\ref{eqn:xQ}).
The retarded Green function $\Gf$ satisfies
\begin{subequations}
    \begin{equation}
    \left[ \p_s + iM(\p_t^2+\w^2) \right] \Gf(t,t';s-s')
    = \delta(t-t')\delta(s-s')
    \label{eqn:Gfa}
    \end{equation}
and the boundary conditions
    \begin{equation}
    \Gf(t,t';s-s')=0\quad\mbox{for}\quad
      t\,(\mbox{or }t')=\tF,\,\tI\;\;\mbox{or}\;\;s<s'.
    \label{eqn:Gfb}
    \end{equation}
\end{subequations}\noindent
Here we have set the initial fictitious time at
$s_0=-\infty$, in order to achieve the long-time limit
automatically.
This choice makes any initial-value dependence of the
solution $\xQf$ irrelevant, for the system is considered to
be in equilibrium already at finite $s$.
Now we solve the Langevin equation
(\ref{eqn:Langevin_model}) to get a recursive relation for
$x(t,s)$
\begin{equation}
x(t,s)=\xclf(t)+\xQf(t,s)
  -ig\int^{\stf}_{\sti}dt'\int^{\infty}_{-\infty}ds'
    \Gf(t,t';s-s')x^3(t',s'),
\label{eqn:recx}
\end{equation}
from which we obtain the perturbation series
\begin{equation}
x=\xclf+\xQf
  -ig\int^{\stf}_{\sti}dt'\int^{\infty}_{-\infty}ds'
      \Gf\left(
        {\xclf}^3+3{\xclf}^2\xQf+3\xclf{\xQf}^2+{\xQf}^3
      \right)% \nonumber\\
+O(g^2).
\label{eqn:expansion}
\end{equation}

\begin{figure}[tb]
%%\figurebox{13.5cm}{2cm}
%\epsfysize=2cm
%\centerline{\epsfbox{expansion.eps}}
\begin{center}
\psbox[scale=0.7]{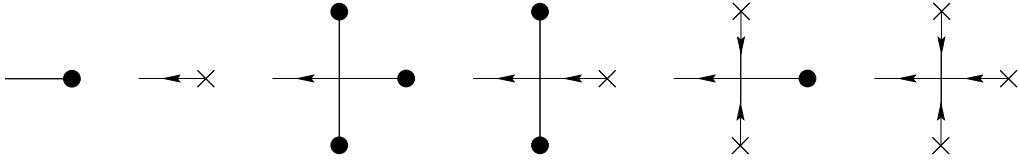}
\end{center}
\caption{%
Graphical representation of the perturbation series of $x$.
}
\label{fig:expansion}
\end{figure}

We represent the series (\ref{eqn:expansion})
diagrammatically as in \reffig{fig:expansion},
where $\xclf$, $\Gf$, and $\eta$ are denoted by a line with
a dot, a line with an arrow, and a cross, respectively, and
at each vertex, a factor $-ig$ and integration variables
$t'$ and $s'$ are attached.
The stochastic average over $\eta$ (\ref{eqn:noise})
combines all the crosses in pairs in all possible ways.
We also have to integrate over the vertex times $t'$ and
$s'$ (see (\ref{eqn:expansion})).

\begin{figure}[tb]
%%\figurebox{7.3cm}{2cm}
%\epsfysize=2cm
%\centerline{\epsfbox{x.eps}}
\begin{center}
\psbox[scale=0.7]{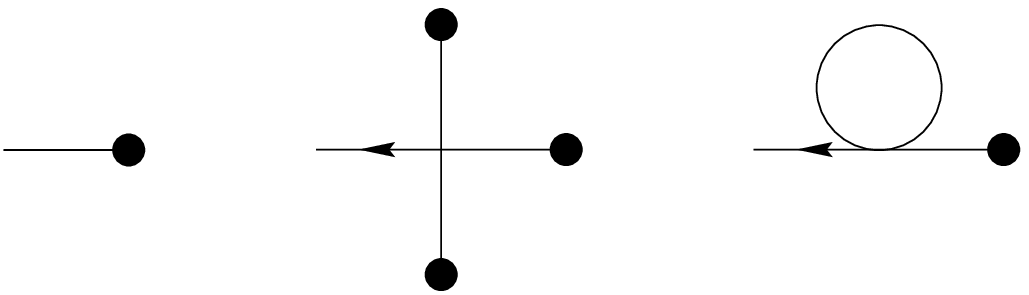}
\end{center}
\caption{%
Stochastic diagrams contributing to $\lb x{\protect\rb}$ up
to $O(g)$.
}
\label{fig:x}
\end{figure}

\begin{figure}[tb]
%%\figurebox{11.6cm}{2cm}
%\epsfysize=2cm
%\centerline{\epsfbox{xx.eps}}
\begin{center}
\psbox[scale=0.7]{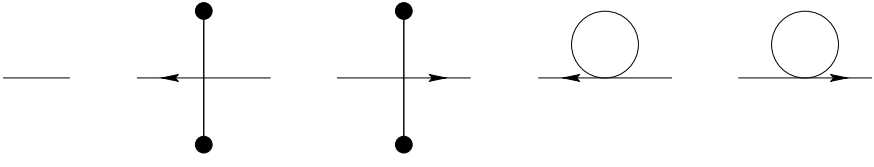}
\end{center}
\caption{%
Connected stochastic diagrams contributing to $\lb
xx{\protect\rb}$ up to $O(g)$.
}
\label{fig:xx}
\end{figure}

According to the formula (\ref{eqn:pti}), we need to
evaluate $\lb x\rb$ in order to get $\lb p(\tI)\rb$, to
which the three diagrams in \reffig{fig:x} contribute up to
$O(g)$,
where a simple line without any arrow or dot denotes the
free two-point correlation function
\begin{eqnarray}
&&\hskip-.4in\Df(t,t';s-s')\nonumber\\
&\equiv&\lb\xQf(t,s)\xQf(t',s')\rb\nonumber\\
\noalign{\vspace{8pt}}
&=&2\int^{\stf}_{\sti}dt_1\int^{\infty}_{-\infty}ds_1
    \Gf(t,t_1;s-s_1)\Gf(t',t_1;s'-s_1)\nonumber\\
&=&\frac{2i}{M\w^2 T}
    \sum_{n=1}^{\infty}
      \frac{1}{\left(\frac{n\pi}{\w T}\right)^2-1}
      \sin\frac{n\pi}{T}(t-\tI)\sin\frac{n\pi}{T}(t'-\tI)
      e^{-iM\w^2\left[
        1-\left( \frac{n\pi}{\w T} \right)^2
      \right]|s-s'|}. \nonumber\\
&&
\label{eqn:Df}
\end{eqnarray}
These three diagrams correspond to the following terms in
$\lb x\rb$
\begin{eqnarray}
&&\hskip-1.0cm\lb x(t,s)\rb\nonumber\\
&=&\xclf(t)
 -ig\int^{\stf}_{\sti}dt'\int^{\infty}_{-\infty}ds'
   \Gf(t,t';s-s')\left(
     {\xclf}^3(t')+3\xclf(t')\Df(t',t';0)
   \right) \nonumber\\
&&\qquad\mbox{}+O(g^2).
\label{eqn:xbc}
\end{eqnarray}
A relation between $\Gf$ and $\Df$ (see the Appendix)
enables us to perform the integration over $s'$ and yields
\begin{equation}
\lb x(t)\rb
=\xclf(t)-ig\int^{\stf}_{\sti}dt' \prop(t,t')
  \left({\xclf}^3(t')+3\xclf(t')\prop(t',t')\right)
+ O(g^2),
\label{eqn:x}
\end{equation}
where $\prop(t,t')\equiv\Df(t,t';0)$ and its explicit form
is\cite{ref:hh-hn}
\begin{equation}
\prop(t,t')=\frac{i}{M\w\sin\w T}
  \sin\w(\tF-\max(t,t'))\sin\w(\min(t,t')-\tI).
\label{eqn:propagator}
\end{equation}
On the other hand, the expectation value $\lb p^2(\tI)\rb$
is calculated from the correlation function $\lb xx\rb$
through the relation (\ref{eqn:ppti}).
At the lowest order, there are three types of connected
diagrams, shown in \reffig{fig:xx},
contributing to $\lb xx\rb$ and we have, after the
integration over $s'$,
\begin{eqnarray}
\lb x(t)x(t')\rb
&=&\lb x(t)\rb\lb x(t')\rb+\prop(t,t')\nonumber\\
&&\quad\mbox{}
    - 3ig\int^{\tF}_{\tI}dt_1\prop(t,t_1)\left(
        {\xclf}^2(t_1)+\prop(t_1,t_1)
      \right)\prop(t_1,t') \nonumber\\
&&\qquad\mbox{}+O(g^2).
\label{eqn:xx}
\end{eqnarray}

We are now in a position to write the explicit form of the
transition amplitude $\la\xF,\tF|\xI,\tI\ra$ for the present
system.
Since the Hamiltonian of this system is given by
\begin{equation}
H=\frac{1}{2M}p^2+\frac{1}{2}M\w^2x^2+\frac{g}{4}x^4,
\label{eqn:Hforx4}
\end{equation}
we have
\begin{equation}
\lb H(\tI)\rb
=\frac{1}{2M}\lb p^2(\tI)\rb+\frac{1}{2}M\w^2\xI^2
+\frac{g}{4}\xI^4,
\label{eqn:exp_Hamiltonian}
\end{equation}
where the last two terms are a consequence of the boundary
conditions (\ref{eqn:boundary}) for $x$.
{}From (\ref{eqn:x}) and (\ref{eqn:xx}) we calculate $\lb
p(\tI)\rb$ and $\lb H(\tI)\rb$ through
(\ref{eqn:exp_Hamiltonian}) and (\ref{eqn:p_configuration}),
and insert them into the formulae (\ref{eqn:formulae}).
In this way, we obtain $\la\xF,\tF|\xI,\tI\ra$ in the
exponential form
\begin{subequations}
\label{eqn:result}
  \begin{equation}
  \la\xF,\tF|\xI,\tI\ra
    =\sqrt{\frac{M\w}{2i\pi\sin\w T}}\,e^{iS_0+iS_1+O(g^2)},
  \end{equation}
where
  \begin{equation}
  S_0=\frac{M\w}{2\sin\w T}\left[
    (\xF^2+\xI^2)\cos\w T-2\xF\xI
  \right]
  \end{equation}
and
  \begin{eqnarray}
  S_1&=&g\frac{3}{32M^2\w^3\sin^2\w T}\left(
          3\w T-3\sin\w T\cos\w T-2\w T\sin^2\w T
        \right) \nonumber\\
     &&\mbox{}
      + ig\frac{3}{16M\w^2\sin^3\w T}\left[
          (\xF^2+\xI^2)\left(
            3\w T\cos\w T-3\sin\w T+\sin^3\w T
          \right)
        \right.\nonumber\\
     &&\makebox[11em]{}
        \left.\mbox{}
          -2\xF\xI\left(
            3\w T-3\sin\w T\cos\w T-2\w T\sin^2\w T
          \right)
        \right]\nonumber\\
     &&\mbox{}
      - g\frac{1}{32\w\sin^4\w T}\left[
          (\xF^4+\xI^4)\left(
            3\w T-3\sin\w T\cos\w T-2\sin^3\w T\cos\w T
          \right)
        \right.\nonumber\\
     &&\makebox[8em]{}
          - 4(\xF^3\xI+\xF\xI^3)\left(
            3\w T\cos\w T-3\sin\w T+\sin^3\w T
          \right)\nonumber\\
     &&\makebox[11em]{}
        \left.\mbox{}
          + 6\xF^2\xI^2\left(
              3\w T-3\sin\w T\cos\w T-2\w T\sin^2\w T
            \right)
        \right].\nonumber\\
  \end{eqnarray}
\end{subequations}\noindent
Notice again that the constant $c$ in
(\ref{eqn:toHamiltonian}) contains enough information to fix
the $T=\tF-\tI$ dependence of $\tilde c$ in
(\ref{eqn:toMomentum}).
On the other hand, the latter can fix the $\xF(\xI)$
dependence of the former.
The remaining factor has been fixed from the normalization
condition at $\tF=\tI$.
The result (\ref{eqn:result}) coincides with the ordinary
perturbative expression for the transition amplitude if it
is expanded as a power series in the coupling constant $g$,
as it should be.

%#################
\section{Summary}
\setcounter{equation}{0}
\label{sec:summary}
In this paper, we have proposed a perturbative treatment of
transition amplitudes within the stochastic quantization
scheme and have applied it to an anharmonic oscillator.
Our treatment is based on the Langevin equation:
We solve it iteratively to get its solution $x$
perturbatively under given boundary conditions.
This enables us to calculate $\lb H(\tI)\rb$ and $\lb
p(\tI)\rb$ as power series in $g$, from which the transition
amplitude is obtained by making use of the formulae
(\ref{eqn:formulae}).
Even though up to the order of interest ($O(g)$, in this
case), they coincides with each other, the expression
(\ref{eqn:result}) is different from the one usually
obtained, for example, in the path-integral formulation:
In this scheme, it is the argument of an exponential
function that is expanded as a power series in $g$.
In this sense, we can say that the present stochastic scheme
systematically incorporates higher-order effects.

It is worth stressing again that the quantity to be
calculated perturbatively in this scheme is not the
amplitude itself, but its logarithm.
This is rooted in the exponential form of the formulae
(\ref{eqn:formulae}).
These formulae are quite remarkable, in the sense that all
quantum fluctuations are put together in one exponent
($\int\lb H(\tI)\rb d\tI$ or $\int\lb p(\tI)\rb d\xI$),
which is contrasted to the usual situation in the
path-integral quantization where fluctuations manifest
themselves as different paths.
Though it is not easy to get an intuitive physical image (if
any) of these exponents, it might be helpful to remind that
a similar relation exists between the ordinary generating
functional and its counterpart for connected diagrams.
The study, along such a line of thought, might lead us to a
deeper understanding of the stochastic scheme itself and
shed new light on the meaning of quantization.
%################

\section*{Acknowledgements}
The authors acknowledge useful and helpful discussions with
Prof.~I.~Ohba and Dr.~Y.~Yamanaka.
This work is partly supported by the Grant-in-Aid for
Scientific Research of the Ministry of Education, Science
and Culture, Japan (\#07804017).

\appendix
%\section{A Relation between $\Gf$ and $\Df$}
%\section{A Relation
%  between $G^{(0)}_{bc}$ and $D^{(0)}_{bc}$}
\section*{Appendix}
\refstepcounter{section}
\setcounter{equation}{0}
\label{app:intG}
We shall prove the relation
\begin{equation}
\int^{\infty}_{-\infty}ds'\Gf(t,t';s-s')=\prop(t,t'),
\label{eqn:intG}
\end{equation}
by means of which we get (\ref{eqn:x}).
Here $\prop$ is defined as
\begin{equation}
\prop(t,t')\equiv\left.\Df(t,t';s-s')\right|_{s=s'},
\label{eqn:proptt'}
\end{equation}
with
\begin{equation}
\Df(t,t';s-s')\equiv\lb\xQf(t,s)\xQf(t',s')\rb.
\label{eqn:DFdef}
\end{equation}
This function has the following explicit form (see
(\ref{eqn:noise}), (\ref{eqn:xQf}) and (\ref{eqn:Gf}))
\begin{eqnarray}
&&\hbox{\hskip-.5in}\Df(t,t';s-s')\nonumber\\
\noalign{\vspace{8pt}}
&=&2\int^{\stf}_{\sti}dt_1\int^{\infty}_{-\infty}ds_1
     \Gf(t,t_1;s-s_1)\Gf(t',t_1;s'-s_1)\nonumber \\
&=&\frac{2i}{M\w^2 T}\sum_{n=1}^{\infty}
     \frac{1}{\left(\frac{n\pi}{\w T}\right)^2-1}
     \sin\frac{n\pi}{T}(t-\tI)\sin\frac{n\pi}{T}(t'-\tI)
     e^{-iM\w^2\left[
       1-\left(\frac{n\pi}{\w T}\right)^2
     \right]|s-s'|}.%\nonumber\\
\label{eqn:expDf}
\end{eqnarray}
It is easy to recognize the existence of a relation between
$\Gf$ in (\ref{eqn:Gf}) and $\Df$ in (\ref{eqn:expDf})
\begin{equation}
\Gf(t,t';s-s')=\theta(s-s')\pdif{s'}\Df(t,t';s-s').
\label{eqn:GfDf}
\end{equation}
Hence the integration in (\ref{eqn:intG}) becomes trivial.
The above relation is nothing but the ``golden rule," as
first observed in \refref{ref:equivalence} and later
interpreted as a realization of the fluctuation dissipation
theorem\cite{ref:SQfd}.

%################

\end{document}